\documentclass[
  aps,
  prl,
  twocolumn,
]{revtex4-2}

\usepackage[utf8]{inputenc}

\usepackage{amsfonts}
\usepackage{booktabs}
\usepackage{microtype}

\usepackage[hidelinks]{hyperref}

\usepackage{amsmath}
\usepackage{amsthm}
\usepackage{dcolumn}
\usepackage{graphics}
\usepackage{graphicx}
\usepackage{subfigure}
\usepackage{amssymb}
\usepackage{graphics}
\usepackage{graphicx}
\usepackage{epstopdf}
\usepackage[svgnames]{xcolor}
\usepackage{subfigure}
\usepackage{url}
\usepackage{soul}
\usepackage{bm}  

\usepackage{siunitx}

\newcommand{\me}[1]{\langle #1 \rangle}

\newcommand\dertt[1]{ \frac{\partial{ #1}}{\partial t} }

\newcommand{\dd}{\text{d}}

\newcommand{\dert}[1]{\frac{\dd #1}{\dd t}}

\newcommand*{\gradient}{\bm{\nabla}}
\newcommand*{\laplacian}{\nabla^2}

\newcommand*{\vort}{\bm{\omega}}  

\newcommand*{\vvec}{\bm{v}}
\newcommand*{\vn}{\vvec_\text{n}}
\newcommand*{\vs}{\vvec_\text{s}}
\newcommand*{\vns}{\vvec_\text{ns}}
\newcommand*{\wn}{\vort_\text{n}}
\newcommand*{\ws}{\vort_\text{s}}
\newcommand*{\rhon}{\rho_\text{n}}
\newcommand*{\rhos}{\rho_\text{s}}
\newcommand*{\nun}{\nu_\text{n}}
\newcommand*{\nus}{\nu_\text{s}}

\newcommand*{\kvec}{\bm{k}}
\newcommand*{\kf}{k_\text{f}}  
\newcommand*{\vf}{v_\text{f}}  
\newcommand*{\kL}{k_\text{L}}  

\newcommand*{\Flux}{\Pi}  

\newcommand*\hypovisc{\mu}

\newcommand*{\Uvec}{\bm{U}}
\newcommand*{\UnVec}{\Uvec_\text{n}}
\newcommand*{\UsVec}{\Uvec_\text{s}}
\newcommand*{\Uns}{U_\text{ns}}  
\newcommand*{\UnsVec}{\Uvec_\text{ns}}
\newcommand*{\UnsF}{\widetilde{U}_\text{ns}}  
\newcommand*{\UnsCrit}{\Uns^*}  
\newcommand*{\UnsFCrit}{\UnsF^*}  

\newcommand*{\vRMS}{v_\text{rms}}
\newcommand*{\vnRMS}{\vRMS^\text{(n)}}

\newcommand*{\fMF}{\bm{f}}  
\newcommand*{\fns}{\fMF_\text{ns}}
\newcommand*{\fMFn}{\fMF_\text{n}}
\newcommand*{\fMFs}{\fMF_\text{s}}

\newcommand*{\forcing}{\bm{\varphi}}
\newcommand*{\forcingN}{\forcing_\text{n}}
\newcommand*{\forcingS}{\forcing_\text{s}}
\newcommand*{\forcingRMS}{\sigma_\text{f}}
\newcommand*{\forcingTime}{t_\text{f}}

\newcommand*{\OmegaNS}{\Omega_\text{ns}}
\newcommand*{\OmegaNSf}{\widetilde{\Omega}_\text{ns}}  

\newcommand*{\En}{E_{\text{n}}}
\newcommand*{\Es}{E_{\text{s}}}
\newcommand*{\Ens}{E_{\text{ns}}}

\newcommand*{\eps}{\varepsilon}

\newcommand*{\epsMF}{\eps_{\text{\tiny MF}}}
\newcommand*{\epsIn}{\mathcal{I}}

\newcommand*\epsSmall{\eps_\nu}     
\newcommand*\epsLarge{\eps_\hypovisc}  

\newcommand*\epsLargeRel{Q_\hypovisc}

\newcommand*{\Rey}{\text{Re}}  
\newcommand*{\ReyN}{\Rey_\text{n}}

\newcommand*{\compidx}{\text{c}}

\newcommand*{\kelvin}{\,\text{K}}

\begin{document}

\title{Counterflow-Induced Inverse Energy Cascade in Three-Dimensional Superfluid Turbulence}
\author{Juan Ignacio Polanco}
\author{Giorgio Krstulovic}
\affiliation{%
  Université Côte d'Azur, Observatoire de la Côte d'Azur, CNRS,
  Laboratoire Lagrange, Boulevard de l'Observatoire CS 34229 - F 06304 NICE Cedex 4, France
}

\begin{abstract}
Finite-temperature quantum turbulence is often described in terms of two
immiscible fluids that
can flow with a non-zero mean relative velocity. Such out-of-equilibrium
state is known as counterflow superfluid turbulence. We report here the
emergence of a counterflow-induced inverse energy cascade in three-dimensional
superfluid flows by performing extensive numerical simulations of the
Hall-Vinen-Bekarevich-Khalatnikov model.
As the intensity of the mean counterflow is increased, an abrupt transition,
from a fully three-dimensional
turbulent flow to a quasi two-dimensional system exhibiting a split
cascade, is observed. The findings of this work could motivate new
experimental settings to study quasi two-dimensional superfluid turbulence in
the bulk of three dimensional experiments.
They might also find applications
beyond superfluids in systems described by more than one fluid component.
\end{abstract}

\maketitle

Turbulence is an out-of-equilibrium state observed in fluids when a
large scale separation exists between the forcing scale, at which the fluid is
stirred, and the dissipation scales where energy is efficiently purged out from
the system. As a result of the inherently non-linear dynamics of fluids, energy
is transferred along scales. Such idea led Richardson to propose his cascade
scenario, where in three-dimensional classical turbulence, energy is transferred
towards small scales in a cascade process~\cite{Frisch1995}. Such a
\emph{direct} cascade, i.e.\ with energy flowing from large to small scales, is
ubiquitous in nature. It
also takes place for instance in magnetohydrodynamic turbulence (e.g.\ in the solar
wind~\cite{Sahraoui2010}) and in quantum turbulence~\cite{Barenghi2014}.
It was later realized by Kraichnan that, in two dimensions, due to
the conservation of enstrophy (mean vorticity square), a different scenario takes
place~\cite{Kraichnan1967}. Energy flows towards large scales through an \emph{inverse} cascade,
whereas enstrophy flows toward small scales by a direct cascade. Such
scenario has been confirmed experimentally and numerically
(see~\cite{Boffetta2012} and references therein).

More complex systems, such as stratified rotating turbulence,
magnetohydrodynamics with a strong background field and some decimated models of
turbulence, might even present split cascades and transitions, where fluxes
can change direction depending on some external
parameters~\cite{Sen2012,Deusebio2014,Seshasayanan2014,Biferale2012a}.
Similarly, thin layer flows, where one dimension is progressively squeezed,
exhibit an abrupt transition from
three to two dimensional phenomenology~\cite{Celani2010,Musacchio2017}. More
recently, such kind of abrupt transition
has also been reported in numerical simulations of low-temperature superfluid
turbulent flows~\cite{Muller2020}.

Superfluids are peculiar types of fluids characterized by the complete absence of
viscosity at low temperature and the presence of quantized vortices (filaments
with a quantized circulation). At finite temperatures, such fluids are composed
of two immiscible components: a superfluid with no viscosity, and a viscous normal
fluid~\cite{Donnelly1991}. The latter is described by the Navier-Stokes
equations. Theses two fluids are coupled through a mutual friction force which
arises from the scattering of thermal excitations on quantized
vortices~\cite{Vinen1957,Donnelly1991}.
The two-fluid description, originally proposed by Landau, enables the possibility
of a turbulent state with no classical analogous, in which the mean relative
velocity between these two
components is non-zero. Such out-of-equilibrium state is known as \emph{counterflow}
turbulence and is typically produced by imposing a
temperature gradient in a
channel~\cite{Vinen1957,Barenghi2014}. Recent numerical
studies of counterflow turbulence have shown a tendency of the system to develop
large-scale quasi-two-dimensional structures~\cite{Biferale2019a,Polanco2020}.
This observation suggests the possibility of a counterflow-induced inverse
energy cascade in quantum turbulent flows.

In this Letter, we investigate the emergence of a split energy cascade in
counterflow superfluid turbulence using direct numerical simulations of the
coarse-grained Hall-Vinen-Bekarevich-Khalatnikov (HVBK) model.
We show an abrupt transition from an isotropic 3D flow (in the absence of a
mean counterflow) to a quasi-2D flow as the mean counterflow velocity is
increased. In particular, for strong counterflow, we observe at large scales the
Kolmogorov-Kraichnan phenomenology of two-dimensional turbulence. Such a
large-scale manifestation is pure consequence of counterflow turbulence, and can
thus be seen as a macroscopic manifestation of quantum mechanics. 

At scales larger than the mean inter-vortex distance, finite-temperature
superfluid helium can be described by the HVBK equations.
In this framework, the motion of discrete quantum vortices is replaced by
their effective coarse-grained dynamics, represented by a continuous
superfluid velocity field.
The turbulent velocity fluctuations $\vn$ and $\vs$ of the
normal and superfluid components then follow two coupled Navier-Stokes
equations~\cite{Donnelly1991,Roche2009,Boue2015,Biferale2019a},
\begin{align}
  \label{eq:HVBK_first}
  \dertt{\vvec_\compidx} +
  (\Uvec_\compidx + \vvec_\compidx) \cdot \gradient \vvec_\compidx
  &=
  -\frac{\gradient p_\compidx}{\rho_\compidx}
  + \nu_\compidx \laplacian \vvec_\compidx
  + \fMF_\compidx + \forcing_\compidx,
  \\
  \label{eq:HVBK_last}
  \gradient \cdot \vvec_\compidx
  &= 0,
  \qquad \compidx \in \{\text{n}, \text{s}\}
\end{align}
where the subscript $\compidx$ identifies each component. The normal fluid
viscosity is denoted by $\nun$.
The effective superfluid viscosity $\nus$ accounts for energy dissipation due
to physics not resolved by the HVBK equations, including small-scale mutual
friction, quantum vortex reconnections and Kelvin wave
excitation~\cite{Vinen2002,Boue2015}.
The respective densities of the normal and superfluid are $\rhon$ and $\rhos$,
and the total density of the fluid is $\rho = \rhon + \rhos$.
The two fluids are stirred by independent zero-mean 3D Gaussian random forces
$\forcingN$ and $\forcingS$ of equal variance $\forcingRMS^2$.
In this model, a mean counterflow velocity $\UnsVec = \UnVec - \UsVec$ is
imposed by setting the respective mean velocities of each component,
$\UnVec$ and $\UsVec$.
Despite $\UnsVec$ being a mean quantity, it cannot be removed with a Galilean
transformation, unlike a constant mean flow in classical turbulence.
The case of zero mean counterflow is known as \emph{coflow} quantum turbulence.

The mutual friction forces are $\fMFs =
-(\rhon / \rhos) \fMFn=\fns$, where $\fns$ depends on $\vns = \vn - \vs$.
In the simplest HVBK description, this coarse-grained mutual friction force
reads $\fns = \alpha \Omega_0 \vns$, where $\alpha$ is a temperature-dependent
non-dimensional coefficient~\cite{Vinen2002}, and the mutual friction frequency
$\Omega_0$ is related to the density and polarization of quantum vortices.
When vortex lines are randomly oriented, as is the case in coflowing quantum
turbulence, this frequency may be estimated as
$\Omega_0 \approx \sqrt{\me{|\ws|^2}} / 2$~\cite{Lvov2006,Biferale2018},
where $\ws$ is the coarse-grained superfluid vorticity,
and $\me{\cdot}$ is an average over space.
Under strong counterflow, the vortex orientation is anisotropic, and this expression
may underestimate the actual friction.
In this case, a common approach is to take $\Omega_0$ as an external control
parameter~\cite{Khomenko2016,Biferale2019a}.
Unless stated otherwise, the first estimation is used throughout this work.

We numerically solve
Eqs.~(\ref{eq:HVBK_first}--\ref{eq:HVBK_last}) using a standard fully parallelized
pseudo-spectral solver in a cubic periodic box of size $L=2\pi$~\cite{Homann2009}.
For the sake of simplicity, we only consider here the case of superfluid helium at $T = 1.9\kelvin$,
where the two fluid components have similar densities
($\rhos / \rhon = 1.35$) and viscosities ($\nus / \nun = 1.25$)~\cite{Donnelly1998,Vinen2002}.

The total energy per unit volume of the system is
$\rho E = \rhon \En + \rhos \Es$,
where $E_\compidx = \me{|\vvec_\compidx|^2} / 2$ is the turbulent kinetic energy associated to
each component.  We consider the energy spectra
\begin{equation}
  E_\compidx(k) = \frac{1}{2} \sum_{k \leq |\kvec| < k + 1} |\widehat{\vvec}_\compidx(\kvec)|^2
  \quad \text{ for } k \in \mathbb{Z},\compidx \in \{\text{n}, \text{s}\},
  \label{eq:energy_spectrum}
\end{equation}
where $\widehat{\vvec}_\compidx(\kvec)$ is the Fourier transform of
$\vvec_\compidx$, and $\kvec$ its wave vector.
The total energy spectrum is the weighted average
$E(k)=[\rhon \En(k) + \rhos \Es(k)]/\rho$.
It quantifies the scale-by-scale repartition of turbulent kinetic energy of
the fluid as a whole.
The relative velocity spectrum $\Ens(k)$ is defined by replacing
$\widehat{\vvec}_\compidx$ with $\widehat{\vvec}_\text{ns}$.

\begin{figure}[]
  \centering
  \includegraphics{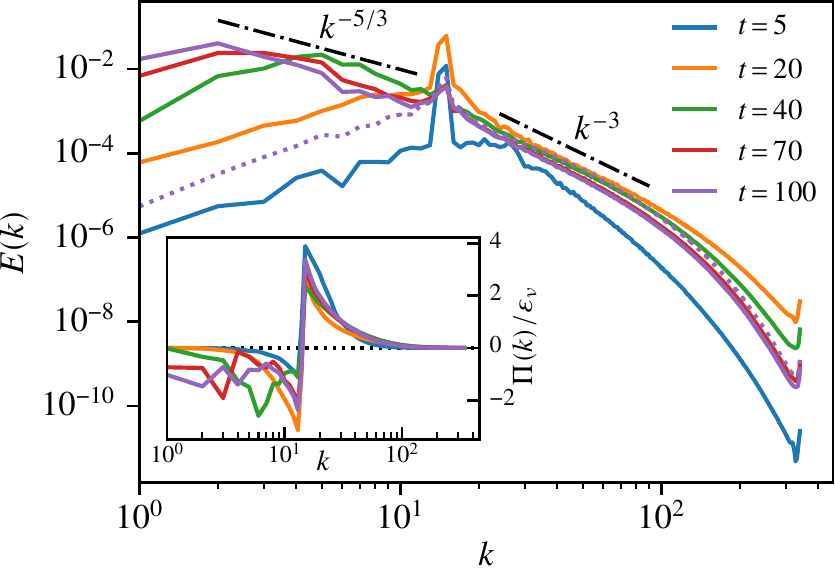}
  \caption{%
    Temporal evolution of the total energy spectrum under
    strong counterflow at $T = \SI{1.9}{\kelvin}$.
    Dotted line: relative velocity spectrum $\Ens(k)$ at the final time.
    In the legend, times are scaled by the forcing time scale
    $\forcingTime = (\kf \forcingRMS)^{-1/2}$.
    Inset: normalized total energy flux.
    At the final time, the normal fluid Reynolds number is $\ReyN = 159$.
  }\label{fig:spectrum_time_evolution}
\end{figure}

A first simulation is performed using $N^3 = 1024^3$ collocation points, with
independent steady 3D forcings $\forcingN$ and $\forcingS$ localized at
the wave number $\kf = 15$.
Initially, the two components have no velocity fluctuations ($\vn = \vs = 0$).
The imposed counterflow velocity, normalized by the forcing velocity
$\vf = \sqrt{\forcingRMS / \kf}$,
is $\UnsF \equiv \Uns / \vf = 40$.
In Fig.~\ref{fig:spectrum_time_evolution}, the time evolution of the total
energy spectrum is shown.
Over time, energy flows from $\kf$ towards both the smallest and the largest
scales of the system, suggesting the formation of a split cascade.
Note that this behavior does not occur in classical three-dimensional
turbulence, where a power-law spectrum $k^n$ with $n\ge1$, usually associated to
thermalized modes, is observed at scales larger than the forcing
one~\cite{Alexakis2019}.
We have observed that both the normal and superfluid energy spectra follow the
same trend as $E(k)$.
In particular, since the two components are locked at large scales, the three
spectra are almost identical for $k < \kf$.
We have also verified that this phenomenon is robust if we turn off the
forcing on the superfluid component, or if both forcings are applied at
different scales.

In the HVBK system, energy is dissipated by the coarse-grained mutual
friction force, by the
kinematic viscosity $\nun$ of the normal fluid, and by the effective viscosity
$\nus$ of the superfluid.
It follows from Eq.~\eqref{eq:HVBK_first} that
\begin{equation}
  \label{eq:energy_balance}
  \dert{E} = -(\epsSmall + \epsMF) + \epsIn,
\end{equation}
where $\rho \epsSmall = \rhon \nun \me{|\wn|^2} + \rhos \nus \me{|\ws|^2}$ is
the small-scale viscous dissipation,
$\rho \epsIn = \rhon \me{\vn \cdot \forcingN} + \rhos \me{\vs \cdot \forcingS}$
is the power injected by the forcing,
and $\epsMF = \OmegaNS \me{|\vns|^2}$ is the dissipation by
mutual friction, with $\OmegaNS = \alpha \rhos \Omega_0 / \rho$.
Note that $\Ens(k)$ is directly related to the mutual friction dissipation as
$\epsMF = 2 \OmegaNS \sum_k \Ens(k)$, and thus characterizes the
scale-by-scale contributions to $\epsMF$. Additionally, as customary in
turbulence~\cite{Frisch1995}, one can define the
energy flux across wave number $k$ as
$
  \Flux_\compidx(k) = \me{\vvec_\compidx^{<k} \cdot [\vvec_\compidx \cdot \gradient \vvec_\compidx]},
$ where $\vvec_\compidx^{<k}$ is the low-pass filtered velocity field
$\vvec_\compidx$ such that $\widehat{\vvec}_\compidx(\kvec)=0$ for $|\kvec| >
k$.
The energy flux $\Flux_\compidx(k)$ quantifies the non-linear transfer of energy from
large scales (such that $|\kvec| \leq k$) to small scales ($|\kvec| > k$).
The total energy flux $\Flux(k)$ is defined as the weighted average of the normal
and superfluid contributions.

The inset of Fig.~\ref{fig:spectrum_time_evolution} shows the energy flux at
different times. Notably, it is negative and relatively flat for $k < \kf$,
indicating the presence of a inverse energy cascade. Conversely, it is positive
for $k > \kf$ indicating also a transfer towards small scales.
The direct cascade builds up rapidly, and has a scaling compatible with
$k^{-3}$.
In classical 2D turbulence, this scaling is associated to a constant enstrophy flux at small scales~\cite{Kraichnan1967,Batchelor1969,Boffetta2012}.
It has also been predicted theoretically in 2D flows with small-scale 3D perturbations~\cite{Moriconi1996}.
Elucidating the origin of such scaling is out of the scope of this Letter, as we focus on the inverse energy cascade.

The build-up of the inverse cascade ($k < \kf$) is slower.
As in 2D turbulence, a Kolmogorov $k^{-5/3}$ spectrum starts to develop at large
scales.
Due to the lack of a large-scale dissipation,
energy accumulates at the largest scales, eventually leading
to the formation of a condensate.

As suggested by the $\Ens(k)$ spectrum in
Fig.~\ref{fig:spectrum_time_evolution} (dotted magenta line), the mutual
friction dissipation is
negligible at scales larger than the forcing, and thus the inverse cascade
dynamics is expected to be similar to that of 2D turbulence.
Namely, energy flows towards the largest scales with negligible loss due to
mutual friction.
This is not the case for the direct cascade, which coexists with a
strong mutual friction dissipation.
Hence, for any given $k > \kf$, a fraction of the energy flows towards smaller
scales, while another part is locally dissipated by mutual friction.
As a result, $\Flux(k)$ monotonically decreases for $k > \kf$, and an inertial
range with a constant energy flux is never observed.

\begin{figure}[]
  \centering
  \includegraphics{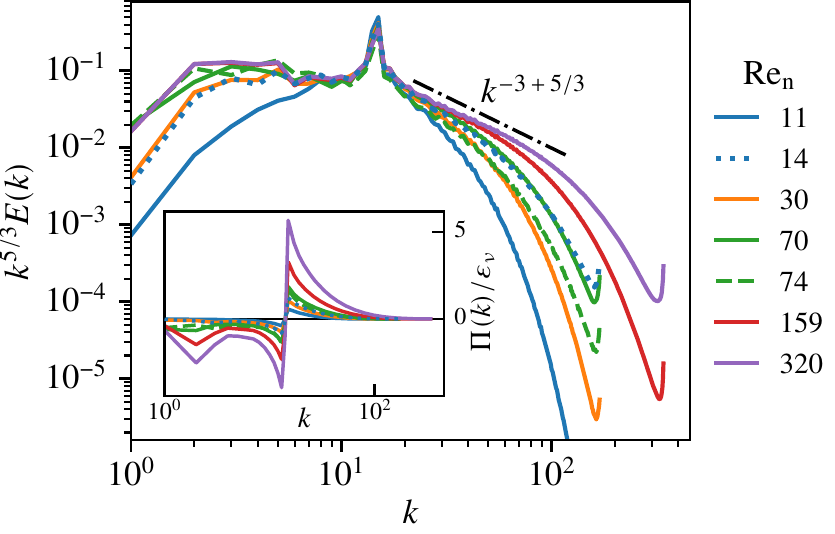}
  \caption{%
    Compensated total energy spectrum under strong counterflow at
    $T = \SI{1.9}{\kelvin}$ for different Reynolds numbers.
    Counterflow values are kept the same, $\UnsF = 40$,
    across all simulations.
      Solid lines, $\nus / \nun = 1.25$ and $\Omega_0^2 = \me{|\ws|^2} / 2$;
      dotted line, viscosity ratio $\nus / \nun = 0.25$ (low $\nus$ case);
      dashed line, externally imposed $\Omega_0$ (high mutual friction case).
    Inset: total energy flux $\Flux(k) / \epsSmall$.
    Spectra and fluxes are averaged over $\Delta t = 30 \, \forcingTime$.
  }\label{fig:spectrum_Reynolds}
\end{figure}

To characterize the effect of the Reynolds number and of the mutual friction
coupling on the split cascade,
we perform simulations at resolutions $N^3 = 512^3$ and $1024^3$ with different
viscosities $\nus$ and $\nun$, while keeping their ratio
$\nus / \nun = 1.25$ constant.
The normal component Reynolds number is $\ReyN = \vnRMS / (\nun \kf)$,
with $\vnRMS$ the standard deviation of $\vn$.
The counterflow velocity is fixed at $\UnsF = 40$.
As shown in Fig.~\ref{fig:spectrum_Reynolds}, the $k^{-5/3}$ scaling of the
inverse cascade is already robust for moderately large Reynolds numbers,
while the direct cascade tends to the $k^{-3}$ scaling at
increasing $\ReyN$.
  Also included is a simulation (dotted lines) with a viscosity ratio
  $\nus / \nun = 0.25$ which also displays an inverse energy cascade.
  This suggests that the choice of effective superfluid viscosity has no
  influence on the large scale dynamics.
Finally, we include a simulation (dashed lines) with an imposed
mutual friction frequency $\Omega_0$ that is 4 times larger than the one
resulting from the $\ws$-based estimate.
The higher coupling between the two components has no apparent influence on the
inverse cascade, while at the small scales, it further suppresses the
velocity fluctuations.
Nevertheless, as confirmed by the energy fluxes (inset of
Fig.~\ref{fig:spectrum_Reynolds}), the double cascade scenario remains mostly
unchanged when the coupling is stronger.

We now proceed to study the transition from the coflowing turbulence with no
inverse cascade to the counterflow-induced double cascade scenario.
For this, we perform a parametric analysis by varying the counterflow velocity
$\Uns$ while setting constant forcing and mutual friction parameters.
The simulations are performed at resolutions $N^3 = 128^3$ and $256^3$.
We now include in Eq.~\eqref{eq:HVBK_first} a large scale dissipation term to
obtain a statistically steady state.
Moreover, to increase the span of the direct and inverse inertial ranges,
the dissipations are strongly localized in wave number space by imposing
hypofriction and hyperviscosity mechanisms~\cite{Boffetta2012}.
These modifications are obtained by replacing the viscous dissipative terms in
Eq.~\eqref{eq:HVBK_first} with
$-[\nu' (-\laplacian)^4 + \hypovisc' (-\laplacian)^{-4}] \vvec_\compidx$.
For simplicity, the two fluid components are given the same values of
$\nu'$ and $\hypovisc'$.
Finally, similar to other studies of transition to 2D
turbulence~\cite{Celani2010,Alexakis2018},
a 2D forcing scheme is introduced, in which the external forces
$\forcing_\compidx$ are orthogonal to the mean counterflow and do not vary in
that direction (i.e.\ $\UnsVec \cdot \forcing_\compidx = \UnsVec \cdot \gradient
\forcing_\compidx = 0$).

The energy balance~\eqref{eq:energy_balance} now writes
$\dd E / \dd t = -(\epsLarge + \epsSmall + \epsMF) + \epsIn$,
with $\epsLarge$ and $\epsSmall$ the large and small-scale dissipations
respectively associated to the hypofriction and hyperviscous terms.
We quantify the strength of the inverse energy cascade by the
relative large-scale dissipation
\begin{equation}
  \epsLargeRel = \frac{\epsLarge}{\epsSmall + \epsLarge}.
\end{equation}
Note that, in contrast to previous studies~\cite{Alexakis2018}, here the
denominator is not the injected power $\epsIn$.
This choice is made because the injected energy is mostly dissipated
locally (in Fourier space) at the forcing scale by mutual friction (as suggested by the $\Ens$
spectrum in Fig.~\ref{fig:spectrum_time_evolution}, peaked at $k = \kf$).

\begin{figure}[]
  \centering
  \includegraphics{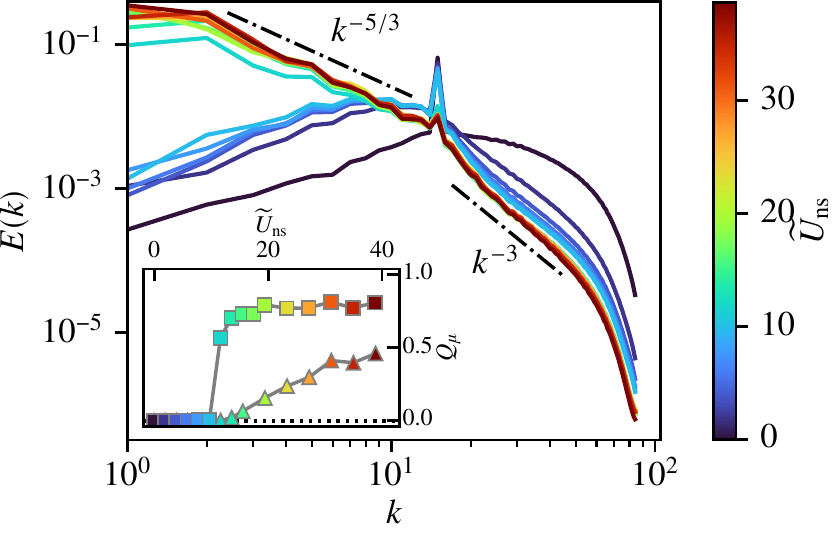}
  \caption{%
    Total energy spectrum for different counterflow velocities $\UnsF$.
    Simulations are performed with a 2D forcing scheme, and include
    hypofriction and hyperviscosity terms (see text for details).
    Inset: relative large-scale dissipation $\epsLargeRel$ as a function of
    counterflow velocity, for 2D (squares) and 3D (triangles) forcing schemes.
    The forcing parameters $(\kf, \forcingRMS)$ are the same across all
    simulations.
  }\label{fig:spectrum_scan_256_hyper}
\end{figure}

The variation of the steady-state energy spectrum $E(k)$ with the imposed
counterflow velocity $\Uns$ is shown in Fig.~\ref{fig:spectrum_scan_256_hyper}
for a set of simulations with $N^3 = 256^3$.
An abrupt transition is observed, from the absence of an inverse cascade at low
$\Uns$, to a double cascade scenario with power laws characteristic of
2D turbulence at large $\Uns$.
In the latter case, the inertial ranges are equivalent to those observed in
higher-resolution simulations (Figs.~\ref{fig:spectrum_time_evolution}
and~\ref{fig:spectrum_Reynolds}), suggesting that the double cascade is not
affected by the dissipation mechanisms at large and small scales.
The dependence of $\epsLargeRel$ with $\UnsF$ (inset of
Fig.~\ref{fig:spectrum_scan_256_hyper}), including for the sake of completeness
the case of a 3D forcing, confirms the appearance of an inverse
energy cascade at a critical value of the counterflow velocity $\UnsFCrit$.
Remarkably, the transition is much more abrupt when the forcing is
two-dimensional than with the original three-dimensional scheme, even though
the value of $\UnsFCrit$ remains almost unchanged.

\begin{figure}[th]
  \centering
  \includegraphics[]{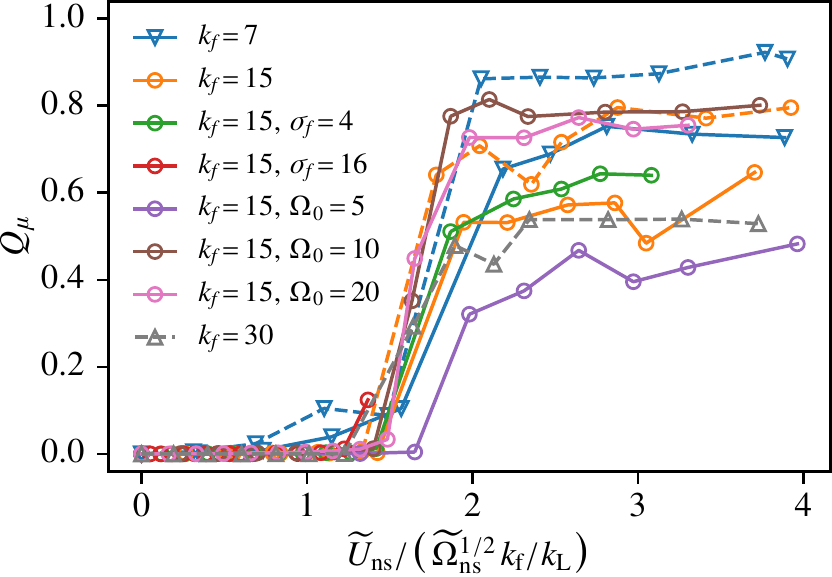}
  \caption{%
    Dissipation ratio as a function of the normalized counterflow velocity,
    for different forcing and mutual friction parameters.
    In all cases a 2D forcing scheme is used.
    Each marker corresponds to a single simulation.
    Resolutions are $N^3 = 128^3$ (solid lines) and $256^3$ (dashed lines), with
    dissipative wave numbers of $k_\eta\approx 100$ and $k_\eta\approx 200$,
    respectively.
    Unless stated otherwise in the legend, the mutual friction frequency
    is $\Omega_0 = \sqrt{\me{|\ws|^2}} / 2$ and the numerical value of the
    forcing intensity is $\forcingRMS = 1$.
  }\label{fig:dissipation_ratio}
\end{figure}

From dimensional analysis, the critical counterflow velocity $\UnsCrit$ can be
expected to depend on the normalized forcing wave number $\kf / \kL$
(where $\kL = 2\pi/L$), and on the
non-dimensional mutual friction intensity
$\OmegaNSf = \OmegaNS / (\kf \forcingRMS)^{1/2}$.
Empirically, from multiple sets of simulations using different forcing and
mutual friction parameters, we find the relation
$\UnsFCrit = C \, \OmegaNSf^{1/2} \kf / \kL$, where $C$ is a
non-dimensional constant.
In terms of dimensional variables, this scaling
becomes $\Uns^* \sim \sqrt{\vf \, \OmegaNS / \kf} \, (\kf / \kL)$.
Note however, that this is an asymptotic formula which assumes that $\OmegaNS$ is
sufficiently large, as for $\OmegaNS=0$ the two fluids are uncoupled and no
transition can be observed.
Figure~\ref{fig:dissipation_ratio} displays the dissipation ratio
$\epsLargeRel$ as a function of the
counterflow velocity scaled according to the above empirical relation, for
different values of the parameters. All simulations invariably display
an abrupt
transition towards a double cascade scenario at nearly the same scaled
counterflow velocity,
which corresponds to a non-dimensional constant $C \approx 1.5$.  Note that we
also present simulations with different values of the small-scale dissipative
wavenumber $k_\eta$, validating the previous scaling. A theoretical explanation,
and further verification of this empirical law, are out of the scope of
this Letter.

We have shown clear evidence of an inverse energy cascade emerging in
finite-temperature quantum turbulence under strong counterflow.
Although described by coarse-grained fluid type equations, this phenomenon can
be seen as a large scale manifestation of quantum mechanics.
Indeed, it originates from the presence
of a counterflow and the coupling between the two fluid components due to mutual
friction, two physical phenomena that arise from quantum mechanical effects.

The appearance of an inverse cascade and the strong bidimensionalization
suggest the possibility of using strong counterflow to produce
(quasi-)two-dimensional turbulent flows in the bulk of three-dimensional
experiments.
Such experiments may be easier to realize than those performed in thin
superfluid helium films~\cite{Donnelly1991,Harris2016,McAuslan2016}. From
Fig.~\ref{fig:spectrum_Reynolds}, we note that the Reynolds numbers needed to
trigger the inverse cascade are relatively low. Therefore, an inverse energy cascade
should be realizable for instance in superfluid experiments with moving or
oscillating objects~\cite{Duda2015,Svancara2017}, provided the experiments are
performed over sufficiently long times, and that some scale separation exists
between the object size and the container.
Such object should move fast enough to ensure a
Reynolds number of order $Re\sim100$ (see Fig.~\ref{fig:spectrum_Reynolds}).
This is not very challenging for current experiments, and is low enough for the
critical counterflow velocity to remain achievable (about 10--15 times the velocity
of the object, see inset of Fig.~\ref{fig:spectrum_scan_256_hyper}).

In this Letter we have only reported the case of superfluid helium at
$T=1.9\kelvin$, where the normal and superfluid densities are similar.
At different temperatures, but
still within the range of applicability of the HVBK model, the situation might be
slightly more complex but the abrupt appearance of an inverse cascade remains
unchanged (data not shown). This will be reported in a future work. Moreover,
note that the large and small scale dissipation mechanisms have no influence on
the emergence of the inverse cascade, making this finding universal.

Finally, we would like to remark that the results of this Letter might find
applications in systems which are not related to superfluid helium, but whose
physics is described by the presence of two or more fluid components.
This is the case for
instance of partially-ionized magnetohydrodynamics occurring in the upper
atmospheres of hot Jupiters and in the interior of Gas Giant
Planets~\cite{Zaqarashvili2011,Benavides2020}. In such systems,
in addition to the induction equation for the magnetic field,
the fluid components are described by
equations strongly resembling the HVBK model.
However, since some components are charged, the components are also coupled to the
magnetic field through the Lorenz force.
It will be then of natural interest, to investigate the consequences of strong
counterflow in the physics of planetary science and other multi-component fluid
systems.

\begin{acknowledgments}
  The authors thank U.~Giuriato and N.~Müller for fruitful discussions.
  This work was supported by the Agence Nationale de la Recherche through the
  project GIANTE ANR-18-CE30-0020-01. GK was also supported by the Simons
  Foundation Collaboration grant ``Wave Turbulence'' (Award ID 651471).
  This work was granted access to the HPC resources of CINES and IDRIS under
  the allocation 2019-A0072A11003 made by GENCI.
  Computations were also carried out at the Mésocentre SIGAMM hosted at the
  Observatoire de la Côte d'Azur.
\end{acknowledgments}

\bibliography{Counterflow}
\bibliographystyle{aipnum4-2}

\end{document}